\documentstyle[12pt]{article}
\newcommand{\beq}{\begin{equation}\label}
\newcommand{\eeq}{\end{equation}}
\newcommand{\p}{\partial}

\newcommand{\ha}{\frac{1}{2}}

\newcommand{\f}{\frac}
\newfont{\testb}{msbm10}
\begin{document}

\title{A finite temperature generalization of Zamolodchikov's C-theorem}

\author{\\Maxim Zabzine\thanks{On leave from Department of Theoretical Physics, St.Petersburg University, Russia}\,\,\thanks{zabzin@vanosf.physto.se}\\
\textit{Institute of Theoretical Physics}\\
\textit{University of Stockholm}\\
\textit{Box 6730}\\
\textit{S-113 85  Stockholm, Sweden}}

\maketitle
.\vskip -12.0cm
\hfill USITP-97-05
\vskip 12.0cm

Abstract: We prove a C-theorem within the  framework  of two dimensional quantum field theories at finite temperature. There exists a function $C(g_{1},g_{2},...,g_{n})$ of coupling constants which is non-increasing along renormalization group  trajectories and non-decreasing along temperature trajectory and  stationary only at the fixed points. The connection between the C-theorem at zero temperature and the C-theorem at finite temperature is discussed. We also consider the thermodynamical aspects of the C-theorem. If we define the C-function in an arbitrary number of dimensions in analogy to the  two dimensional case, we can show that its  behavior  is not universal. The phase transitions destroy the monotonic  properties of the C-function. The proof of the C-theorem is also presented within the framework of the K\"allen-Lehmann spectral representation at finite temperature.      

\newpage

\section{Introduction}

The formalism of standard QFT is suitable for describing observables measured in empty space-time, such as particle interactions in an accelerator. In some interesting physical situations (for example, the early stages of the universe) the environment has a non-negligible matter and radiation density, making the assumptions of standard QFT impracticable. For that reason, we have to look for a new formalism which is closer to thermodynamics. Starting with the basic principles of equilibrium statistical mechanics, the formalism of Feynman's functional integral is used to go from an expression for the time translation operator $exp(-iHt)$ to the partitition function of the grand canonical ensemble $Tr(exp(-\beta H))$ by means of analytical continuation to imaginary time. This formalism is called QFT at finite temperature. On the other hand, the methods of field theory at finite temperature are useful and important for the theory of phase transitions which in turn is connected to spontaneous symmetry breaking in QFT at zero temperature. In this sense QFT at finite temperature is important for undestanding different phases of zero temperature QFT. In this work we  study some general properties of an unitary 2D QFT at finite temperature. We also  discuss the relation between these results and thermodynamics. 

Two dimensional field theories have remarkable properties. One of these properties is that the motion in the space of dimensionless coupling constants under influence of RG group is an ``irreversible'' process. This is the content of the Zamolodchikov's C-theorem which establishes the existence  of a  dimensionless function of the coupling constants  with monotonic properties along RG trajectories \cite{Zam}. We will prove that it is possible to generalize this theorem within the framework of finite temperature 2D QFT. Namely: there exists a function of coupling constants with monotonic properties along RG trajectories and along temperature trajectory. In 2D QFT the RG and temperature flows have similar behavior in the context of the C-function
\beq{0}
\Lambda \f{\p C}{\p \Lambda} = - T \f{\p C}{\p T} \leq 0
\eeq
where $\Lambda$ is the mass scale of the theory and $T$ denotes temperature. This property provides us with a useful tool for studying phases of 2D QFT. We will show that in 2D QFT existence of a C-function is  connected to the triviality of one dimensional statistical mechanics. The finite temperature formulation of C-theorem gives us new insights into the properties of 2D QFT and explains  the difficulties that appear in trying to generalize  Zamolodchikov's theorem to more than two dimensions.

In section \ref{1} we review the different proofs of Zamolodchikov's original theorem. In particular we remind the reader of the  proof where the spectral representation for the two-point function of the stress tensor is used  \cite{spec}. This proof presents a nice physical picture of the RG flow. 

In section \ref{2} we explain what we mean by 2D QFT at finite temperature and also  introduce some notation.

In section \ref{3} we prove  the C-theorem using the finite size approach advocated by Ludwig and Cardy \cite{car}. We note  that there exists a  non-trivial relation between the zero temperature and finite temperature C-theorems. In the finite temperature formulation of the C-theorem the theory becomes  conformally invariant at zero temperature. This IR critical point describes the well-known phase transition at zero temperature for one dimensional statistical system.

In section \ref{4} we disscus a thermodynamical proof of the C-theorem. The monotonic properties of the C-function require  an  additional thermodynamical condition. The standard thermodynamical axioms do not imply this condition and this shows that the C-theorem is not a  universal property of the QFT in arbitrary number of dimensions. The phase transitions destroy the monotonic behavior of the C-function.

In section \ref{5} we generalize the K\"allen-Lehmann spectral representation to the two-point correlator at finite temperature. 

In section \ref{6} we use the this  representation of two-point function of the stress tensor to  prove the finite temperature C-theorem. 

\section{The C-theorem at zero temperature}\label{1}

Irreversibility of the RG flow is proved in two dimensions by Zamolodchikov \cite{Zam}. It says that there exists a function $C(g_{1}, g_{2},..., g_{n})$, which is monotonically decreasing along the RG flow
\beq{a1}
\Lambda \f{dC}{d \Lambda} = - \beta_{i}(g) \f{\p}{\p g_{i}} C(g) \leq 0
\eeq
and stationary for conformally invariant theories
\beq{a2}
\beta_{i}(g)=0 \,\,\,\,\Longleftrightarrow \,\,\,\, \f{\p C}{\p g_{i}} =0
\eeq
where it takes the value of the Virasoro central charge. Zamolodchikov's proof is very simple and uses the conditions of renormalizability, positivity, the translational and rotational symmetries (the Poincar\'e symmetry), and certain special properties of a 2D conformal field theory.

Let us sketch it. Consider the stress tensor $\cal{T}_{\mu \nu}$ and introduce the notation $\Theta \equiv \cal{T}_{\mu}^{\mu}$ and $\cal{T} \equiv \cal{T}$$_{zz}$, where $z = x_{1} + i x_{2}$ are complex coordinates. The correlators of the stress tensors have no anomalous dimensions so they  can be parametrized as follows,
$$<\begin{cal}
T
\end{cal}(z, \bar{z}) \begin{cal}
T
\end{cal}(0, 0)> = \f{F(z \bar{z} \Lambda)}{z^{4}}\,\,\,\,\,\,\,\, <\Theta(z, \bar{z})\begin{cal}
 T
\end{cal}(0, 0)> = \f{G(z \bar{z} \Lambda)}{z^{3} \bar{z}}$$
\beq{a3}
< \Theta(z, \bar{z}) \Theta(0, 0) > = \f{H(z \bar{z} \Lambda)}{z^{2} \bar{z}^{2}}
\eeq
The Poincar\'e invariance\footnote{By Poincar\'e invariance we mean the rotational and translational invariance in euclidian QFT.} requires conservation of the stress tensor, which gives different relations between the scalar functions $F$, $G$, $H$. Let us introduce the quantity
\beq{a4}
C = 2 ( F - \ha G - \f{3}{16} H)
\eeq
which satisfies
\beq{a5}
\dot{C} = - \f{3}{4} H , \,\,\,\,\,\,\,\,\,\,\,\dot{C} \equiv  z \bar{z} \f{d C}{d(z \bar{z})}
\eeq
Unitarity of the theory implies that $H$ is positive definite. Then  $C$ reduces to the central charge at the fixed point ($\Theta = 0$). Introducing the renormalization point we obtain the C-function as a function of coupling constants
\beq{a6}
C(g_{1}, g_{2}, ..., g_{n}) = C|_{|z|=1}
\eeq

An alternative proof of the C-theorem was given by Friedan \cite{fri}, using the K\"allen-Lehmann spectral representation of the correlator of two stress tensors
\beq{a7}
<\begin{cal}
T
\end{cal}_{\mu \nu}(x) \begin{cal}
T
\end{cal}_{\rho \sigma}(0)> = \f{\pi}{3} \int\limits_{0}^{\infty} d\mu c(\mu) \int \f{d^{2}p}{(2\pi)^{2}} e^{ipx} \f{(g_{\mu \nu} p^{2} - p_{\mu} p_{\nu})(g_{\rho \sigma} p^{2} - p_{\rho} p_{\sigma})}{p^{2} + \mu^{2}}
\eeq
In two dimensions there is only one Lorentz invariant object with four indices  which can be  constructed from $p_{\mu}$ and which is compatible with conservation of the stress tensor. Consequently in the spectral representation we have only one unknown scalar function of the intermediate mass scale $\mu$ namely  the spectral density $c(\mu)$. This density also depends on the mass scale $\Lambda$ of the theory, as well as on the  dimensionless coupling constants. The proof of the C-theorem goes on by establishing the properties of $c(\mu)$ \cite{spec}. Unitarity of the theory gives $c(\mu) \geq 0$. The form of $c(\mu)$  in a scale invariant theory is completly fixed by its dimensionality. In two dimensions the scale invariance is equivalent to conformal invariance so this argument provides a  good definition of the spectral density.  Therefore in theories with no  scale invariance the general form of $c(\mu)$ is 
\beq{a8}
c(\mu) = c_{0} \delta(\mu) + c_{1}(\mu, \Lambda)
\eeq
where $c_{0}$ is the central charge of the  Virasoro algebra and $c_{1}$ describes the behavior  away from the critical point $\mu = 0$ and depends on the mass scale $\Lambda$ of the theory. Since the correlators of the stress tensor cannot develop anomalous dimensions, the behavior of $c(\mu)d\mu$ under the RG flow is simply given by dimensional analysis
\beq{a9}
c_{\lambda}(\mu, \Lambda) d\mu = c(\lambda \mu, \Lambda) \lambda d\mu = c(\mu, \Lambda/\lambda)d\mu
\eeq
where we disregard the $\delta(\mu)$ term. For further detailes of the proof we refer the reader to \cite{spec}. 

Let us briefly discuss the connection between Zamolodchikov's and Friedan's proofs. Zamolodchikov's C-function can be obtained by integrating the density $c(\mu)$ with smearing functions $f$ with the following properties: $f > 0$, $f(0) = 1$, $f(\mu)$ decreases exponentially as $\mu \rightarrow{\infty}$ and $\mu d f/d\mu \leq 0$
\beq{a10}
C(g(\Lambda)) = \int d\mu c(\mu) f(\mu) = \int d\mu c_{1}(\mu, \Lambda) f(\mu) + c_{0}
\eeq
Then the derivative along the RG flow is
\beq{a11}
-\beta_{i}(g) \f{\p}{\p g_{i}} C = \Lambda \f{\p}{\p \Lambda} C = \int d \mu c_{1}(\mu, \Lambda) \mu \f{d}{d\mu} f(\mu) \leq 0
\eeq
where we have used (\ref{a9}) and integrated by parts. Zamolodchikov's choice corresponds to
\beq{a12}
\mu \f{d f}{d \mu} = - \f{\pi}{2} \mu^{4} G(|x|=1, \mu)
\eeq
where $G$ is the propogator of a free scalar particle. In the next sections we will use the ideas of these poofs.

\section{Finite temperature 2D QFT}\label{2}

We will consider the grand canonical ensemble of particles which are described by 2D QFT. In the grand canonical ensemble, the isolated system can exchange particles and energy with a reservoir. In this ensemble, the temperature $T$, the volume $V$ and the chemical potential $\mu_{i}$ are fixed variabeles. The grand canonical partitition function is 
\beq{b1}
Z = Tr e^{[-\beta(H - \mu_{i} N_{i})]}\,\,\,\,\,\,\,\,\,\beta = \f{1}{T}
\eeq
where $H$ is the Hamiltonian of 2D QFT and $N_{i}$ are a set of conserved number operators. In our further considerations we study the theory without conserved charges (if the system admits some conserved charge, then we have to make the replacement
\beq{b2}
H(\pi, \phi) \rightarrow \tilde{H}(\pi, \phi) = H(\pi, \phi) - \mu N(\pi, \phi)
\eeq
where $\phi$ is the field and $\pi$ is the conjugate momentum). Using the methods of functional integrals \cite{kap} we can rewrite (\ref{b1}) as 
\beq{b3}
Z = N' \int\limits_{\begin{tiny}
              \begin{array}{lcr} 
              \it{periodic}\\
              \it{antiperiodic}
              \end{array}
             \end{tiny}} D\phi\,\,\, exp( \int\limits_{0}^{\beta} dx_{1} \int dx_{2} \,\,\,\cal{L}) 
\eeq
where $\cal{L}$ is the lagrangian density for 2D QFT and we use the imaginary-time formalism\footnote{Note that we use the definitions and notation of Kapusta \cite{kap}. In another review \cite{rev} of the finite temperature field theory the partitition function is defined as $\int D\phi\,\, exp(-\int_{0}^{\beta}dx_{1}\int dx_{2} \cal{L})$. This definition does not change the general argument presented here, because the monotonic property does not depend on the sign. We can redefine the C-function and obtain the same results.}. We integrate over the space of functions with periodic (antiperiodic) properties
\beq{b4}
\phi( \beta, x_{2}) = \pm \phi( 0, x_{2})
\eeq
which are defined on the strip $\mathbf{R}$ $\times$ $[0, \beta]$. The sign depends on the statistics  of the fields considered, ``$+$'' for bosons and ``$-$'' for fermions. The normalization constant $N'$ is irrelevant for our argument. 

Formally we can consider the finite temperature 2D QFT as a QFT defined on $\mathbf{R} \times S^{1}$. More exactly we may generalize the periodic (antiperiodic) condition (\ref{b5}) to
\beq{b5}
\phi(x_{1}+k\beta, x_{2}) = \pm \phi(x_{1}, x_{2})\,\,\,\,\,\,\,\,\,k \in \mathbf{Z} 
\eeq
and consider QFT on $\mathbf{R}^{2}/\mathbf{\Lambda}$ where $\mathbf{\Lambda}$ is the lattice $\beta \mathbf{Z}$. 

For zero temperature 2D QFT we have the Poincar\'e symmetry $\mathbf{P} = SL(2, \mathbf{R}) \rhd E_{2}$ which is a semidirect product of the Lorentz group $\mathbf SL(2, R)$ and the abelian group of translations $\mathbf E_{2}$. Irreducible representations of the Poincar\'e group are labelled by the eigenvalues of Casimir operators, the spin and the mass. In the finite temperature case, QFT is invariant under the group $\mathbf{P}/\mathbf{\Lambda}$ and the representations of this group can be constructed in the same way as for the standard Poincar\'e group. We have the same Casimir operators and the time-like component of the momentum is quantized in units of $2\pi T$. In other words the time-like component of the momentum  is an element of the dual lattice $\mathbf{\Lambda}^{*}$.

We see that formally finite temperature 2D QFT can be treated as ordinary QFT on a cylinder $\mathbf R \times S^{1}$. The finite temperature 2D QFT can be interpreted as an infinite one dimensional statistical system which has an infinite dimensional phase space and can be obtained from a finite statistical system by  taking the thermodynamical limit
\beq{b6}
\lim_{\begin{tiny}
        \begin{array}{lcr} 
        N \rightarrow \infty\\
        \f{N}{V} = const.
        \end{array}
        \end{tiny}} \int e^{-\beta H} \prod_{i=1}^{N} dq_{i} \prod_{j=1}^{N} dp_{j} = \int D\phi\,\, D\pi\,\, e^{-\beta H(\phi, \pi)}
\eeq
where the number of particles $N$ goes to infinity but the density of particles $N/V$ is finite. We will show how the properties of finite temperature 2D QFT's are connected  with the properties of infinite one dimensional statistical systems. 

We finally define the free energy $F$ of the grand canonical ensemble as
\beq{b6}
F \equiv  - \log Z
\eeq
and the free energy density $\cal{F}$ as
\beq{b7}
\begin{cal}
F
\end{cal} \equiv - \f{1}{V} \log Z
\eeq
where $V$ is the volume of the strip $\mathbf{R}$ $\times$ $[0, \beta]$.

\section{The C-theorem at finite temperature}\label{3}

We are going to use the finite size approach \cite{car}. This is widely used for  solving zero temperature problems, but here we would like to draw the readers attention to the application of these ideas to finite temperature QFT. 

Let us introduce the function $c$
\beq{c1}
c(\lambda, T, a) = \f{6}{\pi T^{2}} [ \begin{cal} 
                                  F 
                                 \end{cal} (\lambda, 0, a) - 
                                 \begin{cal}
                                   F 
                                 \end{cal}  (\lambda, T, a) ]
\eeq   
where $\lambda$ is a set of the coupling constants and $a$ is an UV cut-off. The function $c$ does not need any IR cut-off. The finite-size scaling methods have an  infrared cut-off automatically. The UV divergencies of  (\ref{c1}) are not particular to this construction since UV regularization and renormalization must be introduced at zero temperature, and are then unmodified at finite temperature. 

Let us establish the properties of the  function $c$. By the definitions (\ref{b6}) and (\ref{c1}) $c$ is a dimensionless function. Consider the following infinitestimal transformation
\beq{cb1}
x_{1} \rightarrow x_{1} + \xi_{1} x_{1}, \,\,\,\,\,\,\,\,\, x_{2} \rightarrow x_{2} + \xi_{2} x_{2},\,\,\,\,\,\,\,\,\, |\xi_{1,2}| \ll 1
\eeq
where $x_{1}$ and $x_{2}$ denote the longitudinal and transversal coordinates on the strip $\mathbf{R}$ $\times$ $[0, \beta]$ which we denote as $\mathtt{S}$. The change in the free energy can be represented in terms of the stress tensor
\beq{cb2}
\delta F = - <\delta S> = \f{1}{2\pi}\int\limits_{\mathtt{S}} d^{2}x <\begin{cal}
T
\end{cal}_{11}> \xi_{1} + \f{1}{2\pi} \int\limits_{\mathtt{S}} d^{2}x <\begin{cal}
T
\end{cal}_{22}> \xi_{2}.
\eeq
$S$ denotes the action of our theory. On the other hand the change in $F$ given by (\ref{cb2}) is due to its scale dependence on the inverse temperature $\beta$ and  the UV cut-off (length). So we can read off
\beq{cb3}
\beta \f{\p F}{\p \beta} = \f{1}{2\pi} \int\limits_{\mathtt{S}} d^{2}x <\begin{cal}
T
\end{cal}_{11}> = \f{1}{2\pi} \int\limits_{0}^{\beta} dx_{1} <H>
\eeq
where $H$ is the Hamiltonian of the system
\beq{cb4}
H = \int\limits_{-\infty}^{\infty} dx_{2} \begin{cal}
                                   H
                                  \end{cal} = \int\limits_{-\infty}^{\infty} dx_{2}\begin{cal}
 T
\end{cal}_{11} (x_{1}, x_{2}),   
\eeq
and $\cal{H}$ is the Hamiltonian density. In an unitary theory we have to define the ground state energy as a positive function of the parameters \cite{log}. This assumption implies 
\beq{cb5}
T \f{\p F}{\p T} = - \beta  \f{\p F}{\p \beta} \leq 0 \,\,\,\,  \Longleftrightarrow \,\,\,\, c(\lambda, T, a)V \geq 0   
\eeq
We have thus shown that the function $c$ is positive definite as a consequence of the positivity of the energy of the  ground state.

The free energy density $\cal F$ ($\lambda, T, a$) may be written as
\beq{ca1}
\begin{cal}
F
\end{cal} (\lambda, T, a) = - \f{1}{V} \log [ \int D\phi \,\,exp ( S_{0} + \lambda_{i} \mu^{d_{i}} \int\limits_{\mathtt{S}} d^{2}x O^{i}(x))]
\eeq
where the full action is represented as a deformation of the conformal theory $S_{0}$ by  operators $O^{i}(x)$ of mass dimension $2 - d_{i}$. From (\ref{ca1}) we may read off the expressions
\beq{ca2}
\f{\p c}{\p \lambda_{i}} = \f{6 \mu^{d_{i}}}{\pi T^{2} V} [ \int\limits_{\mathtt{S}} d^{2}x < O^{i}(x) >_{T} - \int\limits_{\mathtt{R}^{2}} d^{2}x < O^{i}(x)>_{0}]
\eeq
If we consider this expression at a nontrivial fixed point\footnote{The trivial fixed point is $\lambda = 0$.} $\lambda_{*}$, then at this point the operators $O^{i}$ have anomalous dimensions $2 - d_{i} - \gamma_{i} (\lambda_{*})$. At the fixed point the theory is scale invariant, so we expect that the expectation values of the operators with nonzero dimension vanish. The calculations of $<O^{i}(x)>_{T}$ and $<O^{i}(x)>_{0}$ require the same UV subtractions. Thus as $<O^{i}(x)>$ have been subtracted in the plane $\mathbf R^{2}$, the expectation value on the strip must vanish at the fixed points \cite{bo}. So at the fixed points the function $c$ is stationary
\beq{ca3}
\f{\p c}{\p \lambda_{i}} = 0
\eeq
In fact we have already proved that the function $c$ is monotonic. The derivatives of $c$ do not change sign between two nearest fixed points. Let us establish that the function $c$ is non-increasing along a RG trajectory. From (\ref{ca2}) and the fact that $\Theta(x)=2\pi \beta_{i}O^{i}(x)$ ($\mu^{d_{i}}$ is included in the definition $O^{i}$) we obtain
\beq{cq1}
-\beta_{i} \f{\p c}{\p \lambda_{i}} = - \f{3}{\pi^{2} T^{2} V} \int\limits_{\mathtt{S}} d^{2}x  <\Theta>_{T}\,\, \leq \,\,0,
\eeq
where $\beta_{i}$ is Gell-Mann-Low functions and $\Theta$ is the trace of the stress tensor. This inequality is the consequence of the positivity condition for the stress tensor \cite{pos}. We choose the UV subtractions is such a way that $<\cal{T}$$_{\mu \nu}>=0$ in the plane. The stress tensor is not a primary operator and its expectation value can not be simply obtained from its value in the plane. So $<\cal{T}_{\mu \nu}>$$_{T}$ on the strip does not vanish and can vanish only at a fixed point.
  
In a scale invariant (conformally invariant) theory the function c is equal to the central charge $c_{0}$ of the corresponding conformal field theory. The finite temperature correction to the free energy density in conformal field theory is given by \cite{car}
\beq{c2}
\begin{cal}
F
\end{cal} (T) =
\begin{cal}
F
\end{cal} (0) - \f{\pi c_{0}}{6} T^{2}
\eeq
The free energy density does not acquire an anomalous dimension under renormalization so $c(\lambda, T, a)$ needs no substractions. This fact implies
\beq{c3}
c(\lambda, T, a) = C(g, \f{\Lambda}{T})
\eeq
where g is a set of renormalizable coupling constants and $\Lambda$ is a renormalization mass scale. In (\ref{c3}) $C(g, \Lambda/T)$ is finite when the cut-off is removed keeping $g$ fixed. The  dependence on $\Lambda$ and $T$  has to be as shown in (\ref{c3}) since $c$ is dimensionless. The function $C$ satisfies the Callan-Symanzik equation
\beq{c4}
(\Lambda \f{\p}{\p \Lambda} + \beta_{i}(g)\f{\p}{\p g_{i}}) C(g, \f{\Lambda}{T}) = 0
\eeq
with Gell-Mann-Low functions
\beq{c5}
\beta_{i}(g) = \Lambda \f{\p g_{i}}{\p \Lambda}
\eeq
These equations can be solved as usual to yield $C=C(g(\Lambda/T))$. All facts which have been proved for $c$ are true for $C$. 

Let us collect all the facts  we have found. Within the framework of finite temperature 2D QFT we can introduce a positiv function of the renormalizable coupling constants $C(g_{1}, g_{2}, ..., g_{n})$ which is non-increasing along a RG trajectory and non-decreasing along a temperature trajectory
\beq{c6}
\Lambda \f{d C}{d \Lambda} = - T \f{d C}{d T} = - \beta_{i}(g) \f{\p C}{\p g_{i}}\,\,\leq\,\,0
\eeq
and which is stationary only at fixed points
\beq{c7}
\f{\p C}{\p g_{i}} = 0 \,\,\,\,\Longleftrightarrow \,\,\,\,\beta_{i}(g) = 0
\eeq
At the critical fixed points, the 2D QFT becomes  conformally invariant and the value of $C$ at these points is the same as the corresponding central charge. 

The proof presented here is not the only possible one within the framework of the finite size approach. This approach is very rich in possibilities. For example, one can represent the C-function  as a two-point function for the stress tensor \cite{mav} and study its properties. 

Let us consider the connection between the C-theorem at finite temperature and Zamolodchikov's original C-theorem at zero temperature. The naive expectation is that if the temperature goes to zero then our finite temperature C-function will become Zamolodchikov's C-function for 2D QFT at zero temperature. But this is not true. At zero temperature the finite temperature 2D QFT becomes  conformally invariant
\beq{c8}
\lim_{T \rightarrow 0} C(g(\f{\Lambda}{T})) = c_{IR}
\eeq
This result can be understood if the finite temperature 2D QFT is interpreted as an infinite one dimesional statistical system. In such a  system the phase transition occurs at zero temperature and there are no other phase transitions at finite temperature \cite{phase}. At the phase transition point  the system has  conformal invariance. So the IR conformal point corresponds to the phase transition at zero temperature. 

There is another, more technical, explanation of the result (\ref{c8}). In zero  temperature QFT  we need an additional mass scale $\Lambda_{0}$ to define the C-function
\beq{c9}
C(g_{1}, g_{2}, ..., g_{n}) = C(g_{1}(\f{\Lambda}{\Lambda_{0}}), g_{2}(\f{\Lambda}{\Lambda_{0}}), ..., g_{n}(\f{\Lambda}{\Lambda_{0}}))
\eeq
This can be seen from simple dimensional analysis. In finite temperature QFT the temperature $T$ plays the role of $\Lambda_{0}$. In  a quantum field theory frame we can understand $\Lambda_{0}$ or $T$ as some kind of infrared cut-off. At the conformal points the free enrgy becomes independent of $T$, which is just an infrared cut-off and can be removed ($T \rightarrow 0$). If we introduce into the  finite temperature prescriptions an additional mass parameter $\Lambda_{0}$, then we  obtain the limit
\beq{c10}
\lim_{T \rightarrow 0} C(g(\f{\Lambda}{\Lambda_{0}}, \f{T}{\Lambda_{0}})) = C(g(\f{\Lambda}{\Lambda_{0}})).
\eeq
But  in finite temperature QFT the additional parameter $\Lambda_{0}$ does not really play any physical role.  

\section{Thermodynamical aspects of the C-theorem}\label{4}

Consider a system of interacting particles in equilibrium where the interactions are described by 2D QFT. We know  how to define the free energy for this system and can define  the other thermodynamical functions as well. We will  re-express the finite temperature C-theorem in terms of the thermodynamical functions. Let us define the C-function in terms of the free energy density as in (\ref{c1}). The positivity of the C-function is equivalent to the positivity of the entropy
\beq{k1}
C(T) \geq 0\,\,\,\,\,\Longleftrightarrow \,\,\,\,\, -\f{\p \begin{cal}
                                                            F
                                                            \end{cal}}{\p T} = \begin{cal}
                                                                               S
                                                                               \end{cal} \geq 0
\eeq
where $\cal{S}$ is the entropy density. Positivity of the entropy is postulated in  thermodynamics, but can be derived in statistical mechanics. As we have shown, this fact is related to the positivity of the energy of the ground state. 

The monotonic property of the C-function (\ref{c6}) can be expressed in terms of the free energy density
\beq{k2}
T\f{d C}{d T} \geq 0 \,\,\,\,\, \Longleftrightarrow \,\,\,\,\, \f{\p \begin{cal}
                                                                     F
                                                                     \end{cal}(T)}{\p T} \leq \f{2}{T}(\begin{cal}
F
\end{cal}(T) - \begin{cal}
                F
               \end{cal}(0))
\eeq
where we use Nernst's theorem $\cal{S}$$(0)=0$. The special inequality (\ref{k2}) for the the free energy density in not a  consequence of the thermodynamical laws. If we represent the free energy density in the standard way in terms of the internal energy density $\cal{E}$ and the entropy density $\cal{S}$
\beq{k3}
\begin{cal}
F
\end{cal}(T) = \begin{cal}
                E
               \end{cal} (T) - T \begin{cal}
                                 S
                                 \end{cal}(T)
\eeq
then the condition (\ref{k2}) can be rewritten as
\beq{k4}
\begin{cal}
S
\end{cal}(T) \leq \f{2}{T}(\begin{cal}
                            E
                           \end{cal}(T) - \begin{cal}
                                           E
                                          \end{cal}(0))
\eeq
The inequality (\ref{k4}) can be proved straightforwardly for one dimensional  statistical systems  and is related to the fact that there are  no phase transitions at nonzero temperature, i.e. no long range order for $T >0$.

In general, a point where  a phase transition at finite temperature $T$ occurs is  defined to be a point where the free energy $F$ is non-analytic. By non-analytical we mean that the function can not be expanded in a Taylor series around that point. So in one dimensional models the free energy is an analytical function of temperature and can be represented as a Taylor series\footnote{In one dimensional models the situation is in fact more sophisticated. We have to evaluate the free energy in every concrete situation. But there exists general theorems (Van Hove's theorems) which show that one dimensional systems with finite range interactions do not exhibit phase transition and the free energy for these systems is an anlytical function of the temperature \cite{one}.}. Let us use the definition of the specific heat at constant volume 
\beq{k5}
C_{V}(T) = T\f{\p \begin{cal}
                   S
                   \end{cal}}{\p T} = \f{\p \begin{cal}
                                                   E
                                             \end{cal}}{\p T}
\eeq
which can be integrated using Nernst's theorem $\cal{S}$$(0)=0$
\beq{k6}
\begin{cal}
 S
\end{cal}(T) = \int\limits_{0}^{T} \f{C_{V}(T')}{T'} dT'
\eeq
\beq{k7}
\begin{cal}
E
\end{cal}(T) - \begin{cal}
                    E
               \end{cal}(0) = \int\limits_{0}^{T} C_{V}(T') dT'
\eeq
Using the assumption of  analyticity of $C_{V}$
\beq{k8}
C_{V}(T) = \sum_{k=1}^{\infty} C_{Vk} T^{k}\,\,\,\,\,\,\,\,\,\,C_{V}(0)=0
\eeq
we can show that our system satisfies the condition (\ref{k4}). The stationarity  of the C-function implies
\beq{s1}
T\f{d C}{d T} = 0 \,\,\,\,\,\Longleftrightarrow\,\,\,\,\,T\f{\p \begin{cal}
                                                                 F
                                                                \end{cal}}{\p T} = 2(\begin{cal}
                                                                                      F
                                                                                     \end{cal}(T) - \begin{cal}
                                                                                                     F
                                                                                                    \end{cal}(0))
\eeq
We can solve the differential equation of the free energy density and obtain
\beq{s2}
\begin{cal}
F
\end{cal}(T) = \begin{cal}
                F
               \end{cal} (0) + A T^{2}
\eeq
where $A$ is a constant of integration. The expression (\ref{s2}) gives  the finite temperature correction to the free energy density in conformal field theory.  

We have demonstrated that the finite temperature C-theorem can be proved within the thermodynamics of infinite one dimensional statistical systems. The monotonicity of the C-function is related to the absence of long range order for $T > 0$ in these systems. 

Let us generalize the definition of the C-function (\ref{c1}) to QFT in an arbitrary number $D$ of dimensions
\beq{k9}
C(T) = \f{6}{\pi T^{D}} [\begin{cal}
                          F
                         \end{cal}(0) - \begin{cal}
                                         F
                                        \end{cal}(T)].
\eeq     
The function $C$ is dimensionless and positive. The monotonic property of $C$ implies the condition
\beq{k10}
\begin{cal}
S
\end{cal}(T) \leq \f{D}{(D-1)T}(\begin{cal}
                                 E
                                \end{cal}(T) - \begin{cal}
                                                 E
                                               \end{cal}(0)).
\eeq
As we have discussed this condition is not a consequence of the thermodynamical laws. The thermodynamics requires that the entropy is bounded from below but not from above. 

In a general situation  the condition (\ref{k10}) can be realised only at low temperature where it is possible that the internal energy density $\cal{E}$ dominates $T\cal{S}$ in the free energy density, and the free energy density $\cal{F}$ may be minimized\footnote{In QFT by minimizing the free energy we can find the ground states (vacua).}  by minimizing $\cal{E}$. At  high temperature, the entropy density $\cal{S}$ always dominates in  the free energy, and the free energy density $\cal{F}$ is minimized by maximizing $\cal{S}$. So at high temperature the condition (\ref{k10}) breaks down as well. If the macroscopic states (the vacua) of the system obtained by these two procedures are different, then we conclude that at least one phase transition has occurred at some intermediate temperature\footnote{This argument is usually called the energy-entropy argument.}.

 In terms of QFT we can say that at high and  low temperatures the quantum system has  different vacuas. At high temperature the spontaneously broken symmetry is restored. So we can expect a generalization of the C-theorem for trivial systems only, which describe just one phase, for example the free QFT. We can introduce the C-function for an arbitrary QFT and expect a ``low temperature'' generalization of Zamolodchikov's C-theorem
\beq{k11}
\Lambda \f{d C}{d \Lambda} = - T \f{d C}{d T} \leq 0,\,\,\,\,\,\, T < T_{c},\,\,\,\Lambda > \Lambda_{c}
\eeq
and phase transitions destroy the monotonic behavior of the  C-function.

We have to remark that condition (\ref{k10}) was considered also in an article by Castro Neto and Fradkin \cite{fradkin}. The authors consider a system in statistical mechanics which can be described on a lattice with some characteristic lenght $a$ (UV cut-off). They study the possibility to define a monotonic C-function in the limit $a\rightarrow 0$ and find (\ref{k10}) to be a sufficient condition for this. We hope that the treatment presented here within the framework of finite temperature QFT clarifies the origin of the condition (\ref{k10}).

\section{The K\"allen-Lehmann spectral representation at finite temperature}\label{5}

In QFT the basic idea of a spectral representation consists in constraining the form of a two-point correlator by enforcing Poincar\'e invariance of the propagating intermediate states \cite{zub}. The propagation of states with an intermediate mass $\mu$ is given by the propagator of a free scalar particle of the same mass. We are going to derive the K\"allen-Lehmann representation for finite temperature 2D QFT.

The first step in this direction is to  find a basis of the Hilbert space of the theory. We consider euclidian QFT defined on $\mathbf R\times S^{1}$, where the radius of $\mathbf S^{1}$ (the inverse temperature) is the  parameter of the theory. The Hilbert space consists of eigenfunctions of the Laplace operator $\bigtriangleup$  
\beq{d1}
\bigtriangleup |p_{n}, \mu > = - \mu^{2} |p_{n}, \mu >,\,\,\,\,\,p_{n}^{2} = - \mu^{2} - (2 \pi n T)^{2}
\eeq
where the time-like component of the momentum is discrete $p^{\nu} = (2\pi n T, p)$. These eigenfunctions form a complete basis 
\beq{d2}
\Psi_{p_{n}, \mu} (x_{1}, x_{2}) = <x | p_{n}, \mu > = \frac{e^{ix_{2}p_{n} + x_{1}\sqrt{\mu^{2}+p_{n}^{2}}}}{(2\sqrt{p_{n}^{2} + \mu^{2}})^{1/2}}.
\eeq
\beq{d3}
\Psi_{p_{n}, \mu} (x_{1} + k \beta, x_{2}) = \Psi_{p_{n}, \mu} (x_{1}, x_{2}),\,\,\,\,\, k \in Z
\eeq
The projection operator on the space of  representations of squared mass $\mu^{2}$ is built as
\beq{d4}
\wp_{\mu^{2}}(T) = \sum^{\infty}_{n= -\infty} |p_{n}, \mu> <p_{n}, \mu |
\eeq
where the projector $\wp_{\mu^{2}}(T)$ is a function of temperature. An important remark is that for different temperatures we find different Hilbert spaces. The sum over all the representations of the Poincar\'e group gives  the identity operator in the Hilbert space with fixed temperature
\beq{d5}
I = \int d \mu^{2} \wp_{\mu^{2}}(T)
\eeq
Let us calculate the propagator of a free scalar particle of mass $\mu$ at temperature $T$. This propagator can be obtained by inserting the projector $\wp_{\mu^{2}}(T)$ into the correlation function
$$G(x_{1}, x_{2}, \mu, T) \equiv <\phi (x_{1}, x_{2}) \phi (0, 0)> = <\phi(x_{1}, x_{2}) \wp_{\mu^{2}}(T) \phi (0, 0)> = $$
\beq{d6}
= T \sum^{\infty}_{n = -\infty} \int \frac{dp}{(2\pi)} \frac{e^{ipx_{2}} e^{i2\pi n T x_{1}}}{p^{2} + (2\pi n T)^{2} + \mu^{2}}
\eeq 
where we use the normalization $<p_{n}, \mu | \phi (x_{1}, x_{2}) |0> = \Psi_{p_{n}, \mu} (x_{1}, x_{2})$. 

Now let us consider the correlator of interacting scalar fields at  temperature $T$  
\beq{d7}
<S(x_{1}, x_{2}) S(0, 0)>_{T} = \int d\mu^{2} < S(x_{1}, x_{2}) \wp_{\mu^{2}}(T) S(0, 0)>_{T}
\eeq
The amplitudes $<S(x_{1}, x_{2})|p_{n}, \mu>$ and $<\phi(x_{1}, x_{2})| p_{n}, \mu>$ transform in the same way under the action of the Poincar\'e group and have the same quantum numbers. Therefore they are equal up to a normalization, which can only depend on physical parameters of our theory (the temperature and the Casimirs of the groups of symmetries). In our case we have
\beq{d8}
<S(x_{1}, x_{2})| p_{n}, \mu>_{T} = N_{S}(\mu^{2}, T) <\phi(x_{1}, x_{2})|p_{n}, \mu>_{T}
\eeq
From (\ref{d7}) and (\ref{d8}) we can obtain the spectral representation of two-point Green function at the temperature $T$
\beq{d9}
<S(x_{1}, x_{2}) S(0, 0)> = \int d\mu^{2} c_{S}(\mu^{2}, T) G(x_{1}, x_{2}, \mu, T)
\eeq
$$c_{S}(\mu^{2}, T) = N^{2}_{S}(\mu^{2}, T)$$
where the function $c_{S}(\mu^{2}, T)$ is called the spectral density. 

\section{The C-theorem and spectral representation at finite temperature}\label{6}

Let us consider the K\"allen-Lehmann spectral representation of the correlator of two stress tensor on the strip $\mathbf{R}$ $\times$ $[0, \beta]$,
$$<\begin{cal}
T
\end{cal}_{\mu \nu}(x) \begin{cal}
T
\end{cal}_{\rho \sigma}(0)>_{T} = \f{\pi}{3} \int\limits_{0}^{\infty} d\mu c(\mu, T) \sum_{n=-\infty}^{\infty} T $$
\beq{e1}
\times \int \f{dp}{2\pi} e^{ipx_{2}} e^{i2\pi n Tx_{1}} \f{(g_{\mu \nu}(p^{2}+(2\pi n T)^{2}) - p_{\mu}p_{\nu})(g_{\rho \sigma}(p^{2}+(2\pi n T)^{2}) - p_{\rho}p_{\sigma})}{p^{2} + (2\pi n T)^{2} + \mu^{2}}
\eeq
where the time-like component of the momentum is $p_{1} = 2\pi n T$ and the space-like component is $p_{2} \equiv  p$. As we have discussed,  in two dimensions there exists only one Lorentz invariant object which is allowed in the spectral representation. Therefore we have only one unknown scalar function namely  the spectral density $c(\mu, T, \Lambda)$ which depends on the intermediate mass scale $\mu$, a mass scale $\Lambda$ of the theory and the temperature $T$, as well as on dimensionless coupling constants. Unitarity of the theory implies that $c(\mu, T, \Lambda) \geq 0$. The function $c(\mu, T)d\mu$ is dimensionless. In a scale invariant theory the form of $c(\mu, T)$ is fixed by its dimesionality
\beq{m1}
c(\mu, T) \sim \delta(\mu)\,\,\,\,\,\,\,\,\,\,\,\,\,\,\,\,c(\mu, T) \sim \delta(T)
\eeq
\beq{m2}
c(\mu, T) \sim \f{1}{\mu}\,\,\,\,\,\,\,\,\,\,\,\,\,\,\,\, c(\mu, T) \sim \f{1}{T}
\eeq
The third and fourth possibilities cause  IR problems at $\mu=0$ or at $T=0$. So in a scale invariant theory the spectral density is 
\beq{m3}
c(\mu, T) = c_{0} \delta(\mu) + c_{0}'\delta(T)
\eeq
When we substitude (\ref{m3}) in the spectral representation for the two-point function of trace of the stress tensor
\beq{m4}
<\Theta(x) \Theta(0)> = 0
\eeq
then the trace annihilates the vacuum and the theory is conformally invariant. The general form of spectral density is
\beq{m5}
c(\mu, T) = c_{0}\delta(\mu) + c_{0}'\delta(T) + \f{1}{T} c_{1}(\f{\mu}{T}, \f{\Lambda}{T})
\eeq
where the last term can be represented as shown in (\ref{m5}). Therefore the spectral density can not develop an anomalous dimension so  the behavior under RG flow is given by dimensional analysis. The last term describes the behavior of spectral density away from the critical points $\mu=0$ and $T=0$. 

The form of  (\ref{m5}) is the only  possible one for spectral density $c(\mu, T, \Lambda)$ in 2D QFT at finite temperature. Considering the limit
\beq{m6}
\lim_{\lambda \rightarrow{0}} \lambda \, d \mu \,\, c(\lambda \mu, \lambda T, \Lambda) = \lim_{\lambda \rightarrow{0}} d \mu \,\, c(\mu, T, \f{\Lambda}{\lambda})
\eeq
we see that the points at $\mu=0$ and at $T=0$ describe the IR fixed point of the theory ($\Lambda \rightarrow \infty$). In the IR limit the massive degrees of freedom effectively become massless and the spectral density is compressed  to delta functions. The IR limit of 2D QFT is described by a conformal field theory.

The limit 
\beq{m7}
\lim_{\lambda \rightarrow{\infty}} \lambda \, d \mu \,\, c(\lambda \mu, \lambda T, \Lambda) = \lim_{\lambda \rightarrow{\infty}} d\mu \,\, c(\mu, T, \f{\Lambda}{\lambda})
\eeq
gives the UV fixed point of the theory ($\Lambda \rightarrow 0$). The short (UV) distance behavior of the two-point function (\ref{e1}) is
\beq{m8}
< \begin{cal}
T
\end{cal}_{z z}(z, \bar{z}) \begin{cal}
T
\end{cal}_{z z}(0, 0)> \sim \f{1}{z^{4}},\,\,\,\,\,\,\,\,\,z \rightarrow 0
\eeq
so the UV fixed point corresponds to a conformal field theory. 

The C-function can be introduced by integrating the smooth part of the spectral density with smearing functions $f(x)$ which has  following properties: $f > 0$,  $f(0) = 1$, $f(x)$ decreases expotentially as $\mu \rightarrow \infty$ and $xdf/dx \leq 0$
\beq{m9}
C(g(\f{\Lambda}{T})) = c_{IR} + \int \f{d\mu}{T} c_{1} (\f{\mu}{T}, \f{\Lambda}{T}) f(\f{\mu}{T}) = c_{IR} + \int dx c_{1}(x, \f{\Lambda}{T}) f(x)
\eeq
Then the derivative along RG flow is
\beq{m10}
-\beta_{i}(g)\f{\p}{\p g_{i}} C = \Lambda \f{\p}{\p \Lambda} C = \int dx c_{1}(x, \f{\Lambda}{T}) x\f{d }{d x}f(x) \leq 0
\eeq
where we have used dimensional arguments and integrated by parts. The other technical details of  the proof can be generalized in a straightforward manner from the standard proof of the zero temperature C-theorem \cite{spec}.

\section{Conclusions}\label{7}

We have considered  an equilibrium system of particles which are described by 2D QFT. This system can be studied within the framework of QFT at  finite temperature. We have proved that for these system there exists a function $C(g_{1}, g_{2}, ..., g_{n})$ of the coupling constants which is non-increasing along renormalization group trajectories and non-decreasing along temperature trajectory and stationary only at the fixed points. This finite temperature generalization of Zamolodchikov's C-theorem gives us the solid ground for treating  quantum theory in the context of thermodynamics. We have shown that the C-theorem is not a  consequence of the thermodynamical laws and is based on a special assumption about analyticity of the free energy. This assumption is fulfilled only in one dimensional statistical systems, where there is no long range order for finite temperature $T > 0$. In one dimensional statistical systems the phase transition occurs only at zero temperature, which is described by an IR conformal theory.

We have presented three proofs of the finite temperature C-theorem. The first proof utilizes the finite size approach, which is  widely used for  studying of two dimensional field theories. This approach is very rich in possibilities and is related to concrete calculations. The definition of the C-function depends on the choice of  renormalization prescriptions. The second proof is presented in the context of the thermodynamics of one dimensional statistical systems. The monotonicity of the C-function implies a special condition for the entropy and internal energy, which can be shown to hold under the assumption of  analyticity of the specific heat. The thermodynamical proof of the C-theorem teaches us that the monotonic behavior of the C-function is not universal and can be expected only at low temperature (low energy) in arbitrary non-two dimensioanl QFT. A phase transition destroys the monotonicity of the C-function. The third proof uses the finite temperature generalization of the K\"allen-Lehmann spectral representation for the two-point function of the stress tensor. In the two dimensonal case there is only one Lorentz invariant which is allowed in the spectral representation. The absence of anomalous dimension for the spectral density means that we can use  dimensonal analysis. The dimesionality of $c(\mu, T)$ completely determines the form and properties  of spectral density.

The first and third proofs show  no trivial relation between the finite temperature C-theorem and Zamolodchikov's C-theorem at zero temperature. We have a phase transition at zero temperature and the theory has conformal invariance at this point. From the point of view of QFT at finite temperature the system becomes  scale invariant at zero temperature.

There are a lot of questions left. We have to understand the role of spontaneous symmetry breaking in two dimensioanl QFT. The finite temperature C-theorem tells us that there are  no phase transitions at temperatures $T > 0$. This means that the broken symmetry can be restored only at zero temperature, so in two dimensional QFT no spontaneous symmetry breaking can occur at all. If this is true, then we have to prove this in a rigorous way. The famous theorem of Coleman \cite{coleman} gives us some insight into these problems. Using the argument, that the scalar field in 2D QFT is dimensionless, Coleman has proved that there is no Goldstone effect in two dimensions. 

Another problem is related to  the generalization of Zamolodchikov's C-theorem to QFT in an arbitrary number of dimensions. The thermodynamical arguments show that it is impossible to define a monotonic function of temperature starting from the free energy (or partitition function). This simply means that there is no naive or straightforward  generalization of Zamolodchikov's theorem. Zamolodchokov's theorem can be generalized as a criteria of the existence of different phases in the quantum system.

\section{Acknowledgements}

I am deeply grateful to Ulf Lindstr\"om for many useful discusions and comments. I would like to thank Parviz Haggi-Mani for help during the preparation of this work. The work was supported by the grant of the Royal Swedish Academy of Sciences.


\begin{thebibliography}{99}
\bibitem{Zam} A.B.Zamolodchikov, JETP Lett. 43 (1986) 731; Sov. J. Nucl. Phys. 46 (1987) 1090.
\bibitem{spec} A.Cappelli, D.Friedan and J.I.Latorre, Nucl.Phys. B352 (1991) 616.
\bibitem{car} A.W.W.Ludwig and J.L.Cardy, Nucl.Phys. B285 (1987) 687.
\bibitem{fri} D.Friedan, lectures at the Nordisk Forskarkurs, Recent developments in quantum field theory, July 1990, Laugarvatn, Iceland, unpublished.
\bibitem{kap} J.I.Kapusta, Finite-temperature field theory (University Press, Cambridge, 1989). 
\bibitem{rev} A.Nieto, Int.J.Mod.Phys., 12 (1997) 1431.\\
M.E.Shaposhnikov, preprint CERN-TH/96-280, hep-ph/9610247.
\bibitem{log} N.N.Bogolubov, A.A.Logunov and I.T.Todorov, Introduction to axiomatic quantum field theory (W.A.Benjamin Inc., Massachusetts, 1975)
\bibitem{bo} D.Boyanovsky and R.Holman, Phys.Rev. D40 (1989) 1964.
\bibitem{pos} J.I.Latorre and H.Osborn, preprint hep-th/9703196.
\bibitem{mav} N.E.Mavromatos and J.L.Miramontes, Phys.Let. B226 (1989) 291.
\bibitem{phase} N. Goldenfeld, Lectures on phase transitions and the renormalization group (Addison-Wesley Publishing Company, 1992).
\bibitem{one} Phase transition and critical phenomena, edited by C.Domb and M.S.Green (Academic Press, London, New York, 1972).
\bibitem{fradkin} A.H.Castro Neto and E.Fradkin, Nucl.Phys. B400 (1993) 525. 
\bibitem{zub} C.Itzykson and J.B.Zuber, Quantum field theory (McGraw-Hill, New York, 1980).
\bibitem{coleman} S.Coleman, Commun.Math.Phys., 31 (9173) 259. 
\end{thebibliography}
\end{document}